\documentclass[prd,twocolumn,aps,showpacs,nofootinbib,nobibnotes,superscriptaddress,preprintnumbers]{revtex4}
\usepackage{epsfig}
\usepackage{graphics}
\usepackage{bm}
\usepackage{color}
\usepackage{dcolumn}   
\usepackage[spanish,english]{babel}
\usepackage{bm}     
\usepackage{bbm}       
\usepackage{amssymb}  
\usepackage{amsmath}
\usepackage{latexsym}
\usepackage{float}
\usepackage{ifthen}
\usepackage{caption,subfig}
\usepackage{enumerate}
\usepackage{url}
\usepackage{caption,subfig}
\usepackage{amsopn}
\usepackage{hyperref}

\usepackage{amsfonts}
\usepackage{multirow}
\usepackage{array}
\usepackage{booktabs}
\usepackage{rotating}

\usepackage{ulem}
\def\clap#1{\hbox to 0pt{\hss#1\hss}}

\def\bea{\begin{eqnarray}}
\def\eea{\end{eqnarray}}
\def\be{\begin{equation}}
\def\ee{\end{equation}}

\begin{document}

\title{Scalar-Vector-Tensor Gravity Theories}

\author{Lavinia Heisenberg} \email{lavinia.heisenberg@eth-its.ethz.ch}
\affiliation{Institute for Theoretical Studies, ETH Zurich, 
 Clausiusstrasse 47, 8092 Zurich, Switzerland}

\date{\today}

\begin{abstract}
We construct the consistent ghost-free covariant scalar-vector-tensor gravity theories with second order equations of motion with derivative interactions. We impose locality, unitarity, Lorentz invariance and pseudo-Riemannian geometry as the fundamental terms. In the tensor sector we require diffeomorphism invariance, whereas we allow the vector sector to be gauge invariant or not. The resulting Lagrangians consist of new genuine couplings among these fields with at most two derivatives per field. They propagate five physical degrees of freedom in the gauge invariant case and six degrees of freedom if the gauge invariance is broken. In the corresponding limit of the free general functions in the Lagrangian, one recovers the generalized Proca theories. These scalar-vector-tensor theories will have important implications for cosmological and astrophysical applications, among which we mention the application to inflation and generation of primordial magnetic fields, new black hole and neutron star solutions, dark matter and dark energy.

\end{abstract}


\maketitle

\section{Introduction}
General Relativity is the unique theory for a massless spin-2 field with two propagating degrees of freedom. Maintaining the Lorentz symmetry, unitarity, locality and a pseudo-Riemannian space-time any modification inevitably introduces new degrees of freedom in the gravity sector. These are typically additional scalar, vector or tensor fields. After specializing the additional ingredients, one can quite generally build the corresponding underlying interactions. 

The simplest extension is based on a scalar field. In theis context, one can construct the most general covariant scalar-tensor theories with second order equations of motion, which uniquely leads to Horndeski interactions \cite{Horndeski:1974wa}. While the resulting theory has derivative and non-minimal couplings to the gravity sector with four arbitrary functions, it gives rise to equations of motion no higher than second order. 

Similarly, one can introduce an additional vector field into the gravity sector and construct in a unique way the most general vector-tensor theories with second order equations of motion. If one imposes gauge invariance on the vector field, only one additional coupling of the vector field to the double dual Riemann tensor is possible \cite{Horndeski:1976gi}. Abandoning gauge invariance allows more general vector-tensor theories, the generalised Proca theories \cite{Heisenberg:2014rta}. Apart from the analogous scalar interactions of the longitudinal mode of the vector field, there are two new genuine purely intrinsic vector interactions with no scalar counterpart. One of them must couple non-minimally to the double dual Riemann tensor \cite{Jimenez:2016isa} (see also \cite{VectorTensorTheories}). These vector-tensor theories have very rich cosmological \cite{VTcosmology} and astrophysical \cite{VTastrophys} implications. In this Letter we aim at constructing the scalar-vector-tensor gravity theories with second order equations of motion, for both the gauge invariant and the gauge broken cases. This will constitute an important step towards unifying the two general classes of Horndeski and generalized Proca theories.  

\section{Scalar-Vector-Tensor Theories}
\subsection{Gauge invariant}
We will construct the consistent ghost-free scalar-vector-tensor theories with derivative interactions between these fields, nonetheless with second order equations of motion, by directly extending the separate constructions of the scalar and vector sectors. We will first assume that the vector field is gauge invariant and the tensor field has the standard diffeomorphism invariance. This constrains the allowed interactions significantly. The consistent scalar-tensor theories are the well known Horndeski theories. Similarly, one can construct consistent vector-tensor theories with second order equations of motion, known as the generalized Proca theories. Constructing the latter, the associated gauge invariance of the vector field is explicitly broken. However, it can be reintroduced in a straightforward manner using the Stueckelberg trick. The generalized Proca interactions can be written as scalar-vector theories with non-trivial couplings between the scalar Stueckelberg field and the gauge field. The Stueckelberg field is restricted to enter only through derivatives and hence has a shift symmetry. After applying the Stueckelberg approach, one obtains shift symmetric Horndeski type interactions for the pure scalar Stueckelberg field sector and genuine new couplings between the scalar field and the vector field, including such of derivative nature. One can then generalize these interactions by explicitly breaking the shift symmetry of the Stueckelberg field. The resulting theories are more general since the relative tuning of the interactions is required only for a subclass of the free functions in the Lagrangian. 

The action of the scalar-vector-tensor theories has the following form
\begin{equation}
\mathcal{S}=\mathcal{S}_{\rm ST}+\mathcal{S}_{\rm SVT},
\end{equation}
where the scalar-tensor interactions in $\mathcal{S}_{\rm ST}$ are the standard Horndeski interactions. We shall represent these interactions as $\mathcal{S}_{\rm ST}=\int d^4x \sqrt{-g}\sum_{i=3}^5\mathcal{L}^i_{\rm ST}$ with
\begin{eqnarray}
\mathcal{L}^3_{\rm ST}&=&G_3[\Pi]\\
\mathcal{L}^4_{\rm ST}&=&G_4R+G_{4,X}\left([\Pi]^2-[\Pi^2]\right)\nonumber\\
\mathcal{L}^5_{\rm ST}&=&G_5G_{\mu\nu}\Pi^{\mu\nu}-\frac{G_{5,X}}{6}\left([\Pi]^3-3[\Pi][\Pi^2]+2[\Pi^3]\right) \nonumber
\end{eqnarray}
with the arbitrary functions $G_{3,4,5}(\pi,X)$ depending on the scalar field $\pi$ and its derivatives $X=-\frac12(\partial\pi)^2$, $\Pi_{\mu\nu}=\nabla_\mu\partial_\nu\pi$, $[\Pi]=\Pi^\mu_\mu$ and furthermore $G_{i,X}=\partial G_i/\partial X$ and $G_{i,\pi}=\partial G_i/\partial \pi$.

The vector field $A_\mu$ shall first be gauge invariant, with its dynamics determined by the gauge invariant field strength $F_{\mu\nu}=\nabla_\mu A_\nu-\nabla_\nu A_\mu$ and its dual $\tilde{F}^{\mu\nu}=\frac12\epsilon^{\mu\nu\alpha\beta}F_{\alpha\beta}$. We can build the corresponding scalar-vector-tensor Lagrangians order by order as $\mathcal{S}_{\rm SVT}=\int d^4x \sqrt{-g}\sum_{i=2}^4\mathcal{L}^i_{\rm SVT}$. 
The second order Lagrangian of the scalar-vector-tensor theories can have the general form $\mathcal{L}^2=f_2(\pi, \partial_\mu\pi, F_{\mu\nu}, \tilde{F}_{\mu\nu})$ \cite{Fleury:2014qfa, Heisenberg:2014rta}. The independent scalar quantities that we can build from the arguments of $f_2$ are $X$, $F=F_{\mu\nu}F^{\mu\nu}$, $\tilde{F}=F_{\mu\nu}\tilde{F}^{\mu\nu}$ and $Y=\partial_\mu\pi \partial_\nu \pi F^{\mu\alpha}F^\nu{}_\alpha$, so that the second order Lagrangian of the scalar-vector-tensor theories can be written as
\begin{equation}
\mathcal{L}^{2}_{\rm SVT}=f_2(\pi,X,F,\tilde{F},Y) \,.
\end{equation}
Note, that the scalar quantity $\partial_\mu\pi \partial_\nu \pi \tilde{F}^{\mu\alpha}\tilde{F}^\nu{}_\alpha$ can be expressed in terms of the above scalar quantities and would thus not be an independent contribution. The same is true for $\partial_\mu\pi \partial_\nu \pi F^{\mu\alpha}\tilde{F}^\nu{}_\alpha$. Clearly, the second order Lagrangian of the Horndeski $\mathcal{L}^2_{\rm ST}=G_2(\pi,X)$ is included in $\mathcal{L}^{2}_{\rm SVT}$. 

We can also construct terms that carry more than one derivative per field. The genuine coupling of the Stueckelberg field with the gauge field with more derivatives acting on the scalar field needs more caution. The interactions will be of the form $\mathcal{M}_3^{\mu\nu}\nabla_\mu\partial_\nu\pi$, where the symmetric rank-2 tensor $\mathcal{M}_3$ needs to be constructed from $\partial_\mu\pi$, $\tilde{F}_{\mu\nu}$ and $g_{\mu\nu}$. If one imposes that higher time derivatives be absent, this restricts $\mathcal{M}^{00}_3$ to not have any other time derivatives than $\dot\pi$. Therefore, the gauge field can only enter into $\mathcal{M}_3$ via $\tilde{F}^{0\alpha}\tilde{F}^0{}_\alpha\sim B^2$. Therefore, the third order Lagrangian of the scalar-vector-tensor interactions arises as 
\begin{equation}
\mathcal{L}^3_{\rm SVT}=\mathcal{M}_3^{\mu\nu}\nabla_\mu\partial_\nu\pi
\end{equation}
where the rank-2 tensor $\mathcal{M}_3$ has to have the form
\begin{equation}
\mathcal{M}^{\mu\nu}_3=\left( f_3(\pi,X)g_{\rho\sigma}+\tilde{f}_3(\pi,X)\partial_\rho\pi\partial_\sigma\pi\right) \tilde{F}^{\mu\rho}\tilde{F}^{\nu\sigma}\,.
\end{equation}
For the interactions to be most general, we broke the shift symmetry of the Stueckelberg field and allowed an explicit $\pi$ dependence in the independent functions $f_3$ and $\tilde{f}_3$. Due to the imposed properties on $\mathcal{M}_3$, these interactions do not give rise to higher than second order derivatives in the equations of motion, thus avoiding instabilities. 

Similarly, we can build the next order interactions with similar properties but quadratic in $\partial^2\pi$. The interaction will be this time of the form 
\begin{equation}
\mathcal{L}^{4}_{\rm SVT}=\left( \mathcal{M}_4^{\mu\nu\alpha\beta}\nabla_\mu\partial_\alpha\pi\nabla_\nu\partial_\beta\pi+f_4(\pi,X)L^{\mu\nu\alpha\beta}F_{\mu\nu}F_{\alpha\beta}\right)
\end{equation}
where $L^{\mu\nu\alpha\beta}$ is the double dual Riemann tensor and the tensor $\mathcal{M}_4$ has to be given by
\begin{equation}
\mathcal{M}^{\mu\nu\alpha\beta}_4=\left( \frac12f_{4,X}+\tilde{f}_4(\pi)\right)\tilde{F}^{\mu\nu}\tilde{F}^{\alpha\beta}\,.
\end{equation}
Note, that in contrast to $\mathcal{L}_3$, for the interactions quadratic in $\partial^2\pi$, we included non-minimal couplings in $\mathcal{L}_4$ in order to maintain that the equations of motions remain of second order. This is needed only for the $X$ dependent part but not for the $\pi$ dependence, hence we can further admit the independent function $\tilde{f}_4(\pi)$. Thus, the general scalar-vector-tensor Lagrangians with gauge invariance are constructed as $\mathcal{S}=\int d^4x \sqrt{-g}\left(\sum_{i=3}^5\mathcal{L}^i_{\rm ST}+\sum_{i=2}^4\mathcal{L}^i_{\rm SVT}\right)$ with the purely scalar Horndeski sector $\mathcal{L}^i_{\rm ST}$ and the genuine scalar-vector-tensor interactions $\mathcal{L}^i_{\rm SVT}$ as
\begin{eqnarray}\label{genLagrangianSVT}
\mathcal{L}^2_{\rm SVT}&=&f_2(\pi,X,F,\tilde{F},Y) \\
\mathcal{L}^3_{\rm SVT}&=&\mathcal{M}_3^{\mu\nu}\nabla_\mu\partial_\nu\pi \nonumber\\
\mathcal{L}^{4}_{\rm SVT}&=& \mathcal{M}_4^{\mu\nu\alpha\beta}\nabla_\mu\partial_\alpha\pi\nabla_\nu\partial_\beta\pi+f_4(\pi,X)L^{\mu\nu\alpha\beta}F_{\mu\nu}F_{\alpha\beta}\nonumber\,.
\end{eqnarray}
By construction, these theories contain five propagating degrees of freedom. Note, that the pure gauge invariant vector-tensor theories are recovered in the limit of a constant scalar field. For $f_4=\text{constant}$ we recover the standard Horndeski vector interactions $\sqrt{-g}L^{\mu\nu\alpha\beta}F_{\mu\nu}F_{\alpha\beta}$. 

\subsection{Broken gauge invariance}
\label{Nogauge}
In the previous section we considered the interactions of the scalar-vector-tensor theories, with the vector sector being $U(1)$ gauge invariant. More general interactions can be constructed if this restriction is lifted. Introducing $S_{\mu\nu}=\nabla_\mu A_\nu+\nabla_\nu A_\mu$ together with the effective metric
 \begin{eqnarray}
&&\mathcal{G}^{\mu\nu}_{f_{nj}} = f_{n1}(\pi,X_i)g_{\mu\nu}+f_{n2}(\pi,X_i)\partial_\mu\pi\partial_\nu\pi \nonumber\\
&&+f_{n3}(\pi,X_i)A_\mu A_\nu+f_{n4}(\pi,X_i)A_\mu \partial_\nu\pi
\end{eqnarray}
the genuine scalar-vector-tensor interactions without gauge invariance $\mathcal{S}^{ng}_{\rm SVT}=\int d^4x \sqrt{-g}\sum_{i=2}^6\mathcal{L}^{i,ng}_{\rm SVT}$ generalise to the following Lagrangians
\begin{eqnarray}\label{genLagrangianSVTnoGauge}
\mathcal{L}^{2,ng}_{\rm SVT}&=&f_2(\pi,X_1,X_2,X_3,F,\tilde{F},Y_1,Y_2,Y_3) \\
\mathcal{L}^{3,ng}_{\rm SVT}&=&\Big(f_{31}(\pi,X_3)g^{\mu\nu}+f_{32}(\pi,X_3)A^{\mu}A^{\nu}\Big) S_{\mu\nu}\nonumber\\
\mathcal L^{4,ng}_{\rm SVT} & = & f_{4}(\pi,X_3)R+f_{4,X_3}S_2 \nonumber\\
\mathcal L^{5,ng}_{\rm SVT}& = & \frac{f_5(\pi,X_3)}{2}G^{\mu\nu} S_{\mu\nu}-\frac{f_{5,X_3}}{6}S_3\nonumber\\
&+&\mathcal{M}_5^{\mu\nu}\nabla_\mu\partial_\nu\pi+\mathcal{N}_5^{\mu\nu}S_{\mu\nu} \nonumber\\
\mathcal{L}^{6,ng}_{\rm SVT}&=&f_6(\pi,X_1)L^{\mu\nu\alpha\beta}F_{\mu\nu}F_{\alpha\beta}+ \mathcal{M}_6^{\mu\nu\alpha\beta}\nabla_\mu\partial_\alpha\pi\nabla_\nu\partial_\beta\pi \nonumber\\
&+&\tilde{f}_6(\pi,X_3)L^{\mu\nu\alpha\beta}F_{\mu\nu}F_{\alpha\beta}+ \mathcal{N}_6^{\mu\nu\alpha\beta}S_{\mu\alpha}S_{\nu\beta}\nonumber\,.
\end{eqnarray}
where $i=1,2,3$ and $j=1,2,3,4$ with the functions $X_1=-\frac12(\partial\pi)^2$, $X_2=-\frac12\partial_\mu\pi A^\mu$, $X_3=-\frac12A_\mu A^\mu$ and $Y_1=\partial_\mu\pi \partial_\nu \pi F^{\mu\alpha}F^\nu{}_\alpha$, $Y_2=\partial_\mu\pi A_\nu F^{\mu\alpha}F^\nu{}_\alpha$, $Y_3=A_\mu A_\nu F^{\mu\alpha}F^\nu{}_\alpha$, respectively. Furthermore, we introduced 
the shortcut notations $S_2=([S]^2-[S^2])/4$ and $S_3=( [S]^3-3[S][S^2] +2[S^3])/8$. The rank-2 tensor $\mathcal{M}^{\mu\nu}_5$ is given by
\begin{eqnarray}
&&\mathcal{M}^{\mu\nu}_5= \mathcal{G}_{\rho\sigma}^{h_{5j}} \tilde{F}^{\mu\rho}\tilde{F}^{\nu\sigma}\,.
\end{eqnarray}
with the four arbitrary functions $h_{5j}$. Similarly, the tensor $\mathcal{N}^{\mu\nu}_5$ has exactly the same form, with the functions $h_{5j}$ replaced by independent functions $\tilde{h}_{5j}$\footnote{For an arbitrary background the dependence of the functions $h_{5j}$ and $\tilde{h}_{5j}$ has to be adjusted respectively, such that either the $X_1$ or the $X_3$ dependence shows up dominantly in order to maintain the temporal component of the vector field non-dynamical. The analysis for such a background will be performed in a future work.}. The tensors $\mathcal{M}_6^{\mu\nu\alpha\beta}$ and $\mathcal{N}^{\mu\nu\alpha\beta}_6$ in the sixth order Lagrangian are given by
\begin{eqnarray}
&&\mathcal{M}^{\mu\nu\alpha\beta}_6=2f_{6,X_1}\tilde{F}^{\mu\nu}\tilde{F}^{\alpha\beta}\nonumber\\
&&\mathcal{N}^{\mu\nu\alpha\beta}_6=\frac12\tilde{f}_{6,X_3}\tilde{F}^{\mu\nu}\tilde{F}^{\alpha\beta}\,.
\end{eqnarray}
Since the gauge invariance is now broken, six degrees of freedom will propagate for these interactions. It is worth to mention at this stage, that one can construct these scalar-vector-tensor theories in an analogous way by taking the decoupling limit of the multi-Proca theories. Note, that the interactions in $\mathcal L^{4,5,6,ng}_{\rm SVT}$ can be further generalised by performing the disformal transformation
\begin{equation}
g^{\mu\nu}\to\mathcal{G}^{\mu\nu}_{f_{nj}}\,,
\end{equation}
which will introduce beyond scalar-vector-tensor theories among other things. For instance in $S_2$ this replacement with $S_{\mu\nu}S_{\alpha\beta}\left( \mathcal{G}^{\mu\alpha}_{f_{nj}} \mathcal{G}^{\nu\beta}_{f_{nj}}-\mathcal{G}^{\mu\nu}_{f_{nj}} \mathcal{G}^{\alpha\beta}_{f_{nj}}\right)$ will require the corresponding relative tuning with $f_{nj}(\pi,X_i)R$.

In the previous and in this section we constructed the scalar-vector-tensor theories with and without gauge invariance, respectively. These scalar-vector-tensor theories will have very important and rich applications to cosmology, specially to the early universe cosmology and dark matter phenomenology. In the following we will give a few directions that will be worthwhile to explore in detail.

\section{Inflation and the primordial magnetogenesis}
\label{earlyUniverse}
A natural application of these scalar-vector-tensor theories is the early universe cosmology. The implications will be multifaceted, both for the concrete realisation of the inflation scenario as well as for the generation of primordial magnetic fields. The scalar field $\pi$ can play the role of the inflaton field and the gauge field couplings may generate magnetic fields.

In standard electromagnetism the Maxwell kinetic term $\mathcal{L}_{\rm MW}=\frac12(\vec{E}^2-\vec{B}^2)$ does not depend on the scale factor for a cosmological background with an FLRW metric and hence the electromagnetic field does not feel the expansion of the universe. This is due to the conformal invariance of the Maxwell Lagrangian. Therefore, the magnetic field scales as $\vec{B}^2\sim a^{-4}$. Due this conformal invariance, the conformal vacuum is preserved and one can not generate magnetic fields in the early universe. In order to overcome this difficulty, one either has to explicitly break the conformal invariance or make the vector field massive. In this respect many attempts in the literature were based on considering direct couplings of the inflaton field with the gauge field that break the conformal invariance. Widely used are $\mathcal{L}_{\rm CB}=F_1(\pi)F_{\mu\nu}F^{\mu\nu}$ and $\tilde{\mathcal{L}}_{\rm CB}=F_2(\pi)F_{\mu\nu}\tilde{F}^{\mu\nu}$. Note, that the latter breaks also parity if the inflaton field is a real scalar field and does not contribute to the energy momentum tensor. Furthermore, we have $F_{\mu\nu}\tilde{F}^{\mu\nu}=-4\vec{B}\cdot \vec{E}$. Due these couplings to the inflaton field in the functions $F_1$ and $F_2$, the conformal invariance is broken and one can generate magnetic fields in the early universe. See the review \cite{Durrer:2013pga} for more details on these couplings. The Lagrangians $\mathcal{L}_{\rm CB}$ and $\tilde{\mathcal{L}}_{\rm CB}$ are just specific subclasses of our more general Lagrangian in $\mathcal{L}^2_{\rm SVT}$. The scalar-vector-tensor theories constructed in the previous section can offer a similar mechanism for the generation of the magnetic field. For illustrative purposes let us consider the Lagrangian  
\begin{equation}\label{genLagrangian}
\mathcal{L}=\sqrt{-g}\left(\frac{M_{\rm Pl}^2}{2}R+G_2(\pi,X)\right)+\sqrt{-g}\mathcal{L}_{\rm EM}
\end{equation}
where we choose as an example for the electromagnetic part 
\begin{equation}
\mathcal{L}_{\rm EM}=\left(-\frac{1}{4}f_2(\pi,X)F-\tilde{f}_2(\pi,X)Y+\mathcal{L}^3_{\rm SVT}+\mathcal{L}^4_{\rm SVT}\right)\,.
\end{equation}
We will assume the metric to be of the FLRW form $ds^2=a^2(-d\eta^2+d\vec{x}^2)$ and the vector field as $A_\mu=(A_0, A_i+\partial_i A)$, with the gauge choice $A_0=0$ and $\partial^i A_i=0$. The electromagnetic part of the Lagrangian $\mathcal{L}_{\rm EM}$ for this background takes the following form
\begin{equation}
\mathcal{S}_{\rm EM}=\int d^3xd\eta \left( \frac12 P^2(\pi)(A_i')^2 +  \frac12 Q^2(\pi)(\partial_iA_j)^2 \right)
\end{equation}
where the two functions $P$ and $Q$ are given by
\begin{eqnarray}
P&=& f_2+\frac{2\pi'^2\tilde{f}_2}{a^2}-\frac{4a'\pi'f_3}{a^3}+\frac{a'(4f_4a^2+\pi'^2(2\tilde{f}_4+f_{4,X}))}{a^6}\nonumber \\
Q&=& \frac{2a'\pi'}{a^5}\left(-a^2f_3+\tilde{f}_3\pi'^2\right)
+\frac{\left(a^3f_2+2f_3(a'\pi'-a\pi'')\right)}{a^3} \nonumber \\
&+&\frac{2(4a^2f_4(a'^2-aa'')+a'\pi'(a'\pi'-a\pi'')(2\tilde{f}_4+f_{4,X}))}{a^6}
\end{eqnarray}
We can canonically normalize the gauge field $\vec{A}=\vec{\mathcal{A}}/P(\pi)$ and decompose it into mode functions as $\vec{A}=\sum_{\lambda=\pm}\int\frac{d^3k}{(2\pi)^{3/2}}\vec{e}_\lambda(\vec{k})e^{i\vec{k}\dot\vec{x}}(a_\lambda(\vec{k})\vec{\mathcal{A}}+h.c.)$, with the creation and annihilation operators $a_\lambda(\vec{k})$ and $a^\dag_\lambda(\vec{k})$. The equation of motion of the mode functions of the canonically normalised gauge field is given by
\begin{equation}\label{eqEMmodes}
\mathcal{A}''_\lambda+\left( \frac{Q^2k^2}{P^2}-\frac{P''}{P} \right)\mathcal{A}_\lambda=0\,.
\end{equation}
Similarly, we can decompose the electric and magnetic fields and express them in terms of the mode functions of the gauge field $E_\lambda=-\frac{1}{a^2}\left(\frac{\mathcal{A}_\lambda}{P} \right)'$ and $B_\lambda=\frac{\lambda k}{a^2}\frac{\mathcal{A}_\lambda}{P}$. The magnetic energy density is given by
\begin{eqnarray}
\frac{d\langle \rho_B \rangle}{d\ln k} = \sum_\lambda \frac{k^3}{4\pi^2a^4}P^2\Big|\left(\frac{\mathcal{A}_\lambda}{P} \right)'\Big|^2 
\end{eqnarray}
The amplitude of the vector field at the end of inflation can then be estimated as $\delta_B^2=\frac{d\langle \rho_B \rangle}{d\ln k}\Big|_{a_f}$. We have much freedom in the functions $f_{2,3,4}$ and $\tilde{f}_{2,3,4}$ in the expressions for $P$ and $Q$. Given a concrete ansatz for these functions, we can then solve (\ref{eqEMmodes}) for $\mathcal{A}$ and obtain the associated energy density of the magnetic fields. For a specific power law ansatz for a subclass of interactions the generation of large-scale magnetic fields without strong coupling problem was shown in \cite{Tasinato:2014fia}. Here, we have even more freedom in the interactions to go beyond. Since our new interactions enter with more derivatives compared to those considered in \cite{Tasinato:2014fia}, we expect the generation of the magnetic fields to happen during pre-heating.

\section{Black hole and neutron star solutions}
Further crucial implications of these scalar-vector-tensor theories will be new black hole and neutron star solutions with new features coming from the scalar and vector modes, and their interactions. We expect new hairy black hole solutions from a very similar spirit as in scalar-tensor and generalized Proca theories. As an example let us consider the following Lagrangian
\begin{equation}\label{genLagrangian}
\mathcal{L}=\sqrt{-g}\left(\frac{M_{\rm Pl}^2}{2}R+G_2(\pi,X,F)+\mathcal{L}^3_{\rm SVT}+\mathcal{L}^4_{\rm SVT}\right)\,,
\end{equation}
where $\mathcal{L}^3_{\rm SVT}$ and $\mathcal{L}^4_{\rm SVT}$ are the gauge invariant Lagrangians in equation (\ref{genLagrangianSVT}).
We can assume a spherically symmetric background $ds^2=-f(r)dt^2+h^{-1}(r)dr^2+r^2d\Omega^2$ with the field configurations $\pi=\pi(r)$ and $A_\mu=(A_0(r),A_1(r),0,0)$. Note, that the longitudinal mode corresponds to an unphysical gauge mode and does not contribute to the dynamics. The dependence on $\tilde{F}$ in (\ref{genLagrangianSVT}) vanishes and the $Y$ term does not deliver additional independent interactions. The Lagrangians $\mathcal{L}^{3,4}_{\rm SVT}$ simplify in this field configuration to
\begin{eqnarray}
\mathcal{L}^3_{\rm SVT}&=& \frac{2f_3h^2A_0'^2\pi'}{rf}\nonumber\\
\mathcal{L}^4_{\rm SVT}&=&\frac{hA_0'^2(-4f_4(h-1)+h^2\pi'^2(2\tilde{f}_4+f_{4,X}))}{r^2f}
\end{eqnarray}
The contributions of these Lagrangians to the equations of motion for $\pi$, $A_0$, $h$ and $f$ are very different from the corresponding Lagrangians in the generalized Proca theories studied in \cite{VTastrophys}, hence we expect interesting type of new hairy black hole solutions in the presence of these genuine scalar-vector-tensor interactions. One can specifically construct those solutions with $f=h$ and $A_0\ne0$ and $\pi\ne0$ (see \cite{Heisenberg:2018vti} for more detail). 

\section{Dark Energy models}
If one is willing to apply the scalar-vector-tensor theories presented above to dark energy phenomenology, there will be important differences depending on whether one chooses vector fields with our without gauge invariance. In the gauge invariant case for a homogeneous and isotropic background one would need to promote the gauge field to a non-abelian gauge field, hence enhancing the underlying symmetry to the $SU(2)$ group. Restricting the Lagrangians to a subclass with no derivative non-minimal coupling will allow for the right phenomenology. One could for instance consider the Lagrangian $\mathcal{L}=\sqrt{-g}(R+\sum_{i=2}^4\mathcal{L}^{i}_{SVT})$, where $f_4=0$ in $\mathcal{L}^4_{SVT}$ and $F_{\mu\nu}$ and $\tilde{F}_{\mu\nu}$ are promoted to $F^a_{\mu\nu}$ and $\tilde{F}^a_{\mu\nu}$, respectively. The scalar field can then admit a time dependence $\pi(t)$ and the non-abelian field take the field configuration $A_\mu^a=A(t)\delta^a_i$. The Lagrangian with broken gauge invariance will probably be more appealing for dark energy applications (see \cite{Jimenez:2016upj,ExamplesMultiProca}). In the presence of a single Proca field cosmological solutions can be realised with the temporal component of the vector field \cite{VTcosmology}. One can for instance consider the Lagrangian $\mathcal{L}=\sqrt{-g}(R+\sum_{i=2}^6\mathcal{L}^{i,ng}_{SVT})$ with $f_{4,X_i}=0$, $f_5=0$ and $A_{\mu}=(A_0(t),0,0,0)$. One can also promote these Lagrangians to the case of multi-Proca fields by replacing $A_\mu$ by $A_\mu^a$ and realise three different field configurations: purely temporal $A_\mu^a=\phi^a(t)\delta^0_\mu$, triad $A_\mu^a=A(t)\delta^a_\mu$ and the extended triad $A_\mu^a=\phi^a(t)\delta^0_\mu+A(t)\delta^a_\mu$ \cite{Jimenez:2016upj}. For the last two field configurations the Lagrangians would need to be restricted to $\mathcal{L}=\sqrt{-g}(R+\sum_{i=2}^4\mathcal{L}^{i,ng}_{SVT})$ with $f_{4,X_i}=0$ for the right propagation speed of the tensor modes.

\section{Dark Matter applications}
Scalar-vector-tensor theories have been traditionally widely used as alternatives to dark matter particle. Predominant examples are TeVeS \cite{Bekenstein:2004ne} and MOG \cite{Moffat:2005si}. They correspond to relativistic theories that recover the MOND phenomenology on galactic scales. Within the class of scalar-vector-tensor theories we have proposed here, the broken gauge case would be very appealing as an alternative  to dark matter models. The presence of the massive vector field will yield a repulsive Yukawa force. Our scalar-vector-tensor theories can naturally generalise these examples and offer richer phenomenology due to the new non-trivial couplings between the scalar and vector fields. The coupling of the Proca field with the one scalar field in these theories can be extended by the new interactions in $\mathcal{L}^{i,ng}_{\rm SVT}$.

\section{Conclusion}
In this Letter we presented a new class of modified gravity theories based on an additional scalar and vector field on top of the standard tensor field. The resulting scalar-vector-tensor Lagrangians contain derivative interactions among other things, nevertheless give rise to second order equations of motion. We first constructed these theories under the assumption that the vector field carries a $U(1)$ gauge invariance, which restricts the allowed interactions considerably. Next, we relaxed this condition and allowed interactions breaking gauge symmetry. In both cases the resulting theories contain new genuine couplings between these fields with at most two derivatives per field and five degrees of freedom propagating in the case of gauge-invariant interactions and six in the broken gauge case. One can recover the generalised Proca interactions taking the corresponding limit of the free general functions accordingly.
The scalar-vector-tensor theories presented here will have relevant implications for cosmology and astrophysics. We mentioned a few directions for applications to inflation and generation of primordial magnetic fields, new black hole and neutron star solutions, dark matter and dark energy.

\section*{Acknowledgements}
LH would like to thank G. Horndeski, M. Bartelmann, J. Froehlich, S. Tsujikawa and J. Beltran for useful discussions.
LH thanks financial support from Dr.~Max R\"ossler, 
the Walter Haefner Foundation and the ETH Zurich
Foundation.



\begin{thebibliography}{99}

\bibitem{Horndeski:1974wa} 
  G.~W.~Horndeski,
  Int.\ J.\ Theor.\ Phys.\  {\bf 10}, 363 (1974).

\bibitem{Horndeski:1976gi} 
  G.~W.~Horndeski,
  J.\ Math.\ Phys.\  {\bf 17}, 1980 (1976).
  doi:10.1063/1.522837
  
\bibitem{Heisenberg:2014rta} 
  L.~Heisenberg,
  JCAP {\bf 1405}, 015 (2014)
  [arXiv:1402.7026 [hep-th]].

\bibitem{Jimenez:2016isa} 
  J.~Beltran Jimenez and L.~Heisenberg,
  Phys.\ Lett.\ B {\bf 757}, 405 (2016)
  doi:10.1016/j.physletb.2016.04.017
  [arXiv:1602.03410 [hep-th]].



\bibitem{VectorTensorTheories} 
  G.~Tasinato,
  JHEP {\bf 1404}, 067 (2014)
  [arXiv:1402.6450 [hep-th]];
  E.~Allys, P.~Peter and Y.~Rodriguez,
  JCAP {\bf 1602}, no. 02, 004 (2016)
  [arXiv:1511.03101 [hep-th]];
  L.~Heisenberg, R.~Kase and S.~Tsujikawa,
  Phys.\ Lett.\ B {\bf 760}, 617 (2016)
  [arXiv:1605.05565 [hep-th]];
  R.~Kimura, A.~Naruko and D.~Yoshida,
  JCAP {\bf 1701}, no. 01, 002 (2017)
  [arXiv:1608.07066 [gr-qc]].

\bibitem{VTcosmology} 
  A.~De Felice, L.~Heisenberg, R.~Kase, S.~Mukohyama, S.~Tsujikawa and Y.~l.~Zhang,
  JCAP {\bf 1606}, no. 06, 048 (2016)
  [arXiv:1603.05806 [gr-qc]];
  A.~De Felice, L.~Heisenberg, R.~Kase, S.~Mukohyama, S.~Tsujikawa and Y.~l.~Zhang,
  Phys.\ Rev.\ D {\bf 94}, no. 4, 044024 (2016)
  [arXiv:1605.05066 [gr-qc]];
  L.~Heisenberg, R.~Kase and S.~Tsujikawa,
  JCAP {\bf 1611}, no. 11, 008 (2016)
  [arXiv:1607.03175 [gr-qc]];
  A.~de Felice, L.~Heisenberg and S.~Tsujikawa,
  Phys.\ Rev.\ D {\bf 95}, no. 12, 123540 (2017)
  [arXiv:1703.09573 [astro-ph.CO]];
  J.~D.~Barrow, M.~Thorsrud and K.~Yamamoto,
  JHEP {\bf 1302} (2013) 146
  [arXiv:1211.5403 [gr-qc]];
J.~Beltran Jimenez, R.~Durrer, L.~Heisenberg and M.~Thorsrud,
  JCAP {\bf 1310}, 064 (2013)
  [arXiv:1308.1867 [hep-th]].

  
\bibitem{VTastrophys} 
  L.~Heisenberg, R.~Kase, M.~Minamitsuji and S.~Tsujikawa,
  Phys.\ Rev.\ D {\bf 96}, no. 8, 084049 (2017)
  [arXiv:1705.09662 [gr-qc]];
  L.~Heisenberg, R.~Kase, M.~Minamitsuji and S.~Tsujikawa,
  JCAP {\bf 1708}, no. 08, 024 (2017)
  [arXiv:1706.05115 [gr-qc]];
  J.~Chagoya, G.~Niz and G.~Tasinato,
  Class.\ Quant.\ Grav.\  {\bf 33}, no. 17, 175007 (2016)
  [arXiv:1602.08697 [hep-th]];
  M.~Minamitsuji,
  Phys.\ Rev.\ D {\bf 94}, no. 8, 084039 (2016)
  [arXiv:1607.06278 [gr-qc]];
  E.~Babichev, C.~Charmousis and M.~Hassaine,
  JHEP {\bf 1705}, 114 (2017)
  [arXiv:1703.07676 [gr-qc]];
   A.~Cisterna, M.~Hassaine, J.~Oliva and M.~Rinaldi,
  Phys.\ Rev.\ D {\bf 94}, no. 10, 104039 (2016)
  [arXiv:1609.03430 [gr-qc]].

 
\bibitem{Fleury:2014qfa} 
  P.~Fleury, J.~P.~Beltran Almeida, C.~Pitrou and J.~P.~Uzan,
  JCAP {\bf 1411}, no. 11, 043 (2014)
  [arXiv:1406.6254 [hep-th]].

 
 

\bibitem{Durrer:2013pga} 
  R.~Durrer and A.~Neronov,
  Astron.\ Astrophys.\ Rev.\  {\bf 21}, 62 (2013)
  [arXiv:1303.7121 [astro-ph.CO]].

\bibitem{Tasinato:2014fia} 
  G.~Tasinato,
  JCAP {\bf 1503}, 040 (2015)
  [arXiv:1411.2803 [hep-th]].
  
\bibitem{Heisenberg:2018vti}
  L.~Heisenberg and S.~Tsujikawa,
  arXiv:1802.07035 [gr-qc].

\bibitem{Jimenez:2016upj} 
  J.~Beltran Jimenez and L.~Heisenberg,
  Phys.\ Lett.\ B {\bf 770}, 16 (2017)
  doi:10.1016/j.physletb.2017.03.002
  [arXiv:1610.08960 [hep-th]].
  
  \bibitem{ExamplesMultiProca} 
 E.~Allys, P.~Peter and Y.~Rodriguez,
  Phys.\ Rev.\ D {\bf 94}, no. 8, 084041 (2016)
  doi:10.1103/PhysRevD.94.084041
  [arXiv:1609.05870 [hep-th]];
    R.~Emami, S.~Mukohyama, R.~Namba and Y.~l.~Zhang,
  JCAP {\bf 1703}, no. 03, 058 (2017)
  doi:10.1088/1475-7516/2017/03/058
  [arXiv:1612.09581 [hep-th]];
  Y.~Rodríguez and A.~A.~Navarro,
  Phys.\ Dark Univ.\  {\bf 19}, 129 (2018)
  doi:10.1016/j.dark.2018.01.003
  [arXiv:1711.01935 [gr-qc]].

\bibitem{Bekenstein:2004ne} 
  J.~D.~Bekenstein,
  Phys.\ Rev.\ D {\bf 70}, 083509 (2004)
  Erratum: [Phys.\ Rev.\ D {\bf 71}, 069901 (2005)]
  doi:10.1103/PhysRevD.70.083509, 10.1103/PhysRevD.71.069901
  [astro-ph/0403694].

\bibitem{Moffat:2005si} 
  J.~W.~Moffat,
  JCAP {\bf 0603}, 004 (2006)
  doi:10.1088/1475-7516/2006/03/004
  [gr-qc/0506021].


\end{thebibliography}
\end{document}